\documentclass[a4paper,12pt]{article}  
\usepackage{amssymb}
\usepackage{latexsym}
\usepackage[dvips]{graphicx} 
\newcommand{\be}{\begin{equation}}
\newcommand{\ee}{\end{equation}}
\newcommand{\bea}{\begin{eqnarray}}
\newcommand{\nn}{\nonumber}
\newcommand{\eea}{\end{eqnarray}}

\newcommand{\nk}{\noindent}

\begin{document}

\begin{titlepage}
\begin{flushright}
hep-th/0012134\\
UA/NPPS-15-2000
\end{flushright}
\begin{centering}
\vspace{.8in}
{\large {\bf Are Extremal 2D Black Holes 
Really Frozen ?}} \\ 

\vspace{.5in}
{\bf Elias C. Vagenas\footnote{hvagenas@cc.uoa.gr}} \\

\vspace{0.3in}
University of Athens\\ Physics Department\\ 
Nuclear and Particle Physics Section\\ 
Panepistimioupolis, Ilisia GR 157 71\\ Athens, Greece\\ 
\end{centering}

\vspace{1in}
\begin{abstract}
\nk
In the standard methodology for evaluating the Hawking radiation
emanating from a black hole, the background geometry is fixed.
Trying to be more realistic we consider a dynamical geometry for a
two-dimensional charged black hole and we evaluate the Hawking 
radiation as a tunneling process. This modification to the geometry
gives rise to a nonthermal part in the radiation spectrum. 
We exlore the consequences of this new term for the extremal case. 
    
\end{abstract}
\end{titlepage}

\newpage

\baselineskip=18pt
\section*{Introduction} 
\hspace{0.8cm}
There has been much work on the classically forbidden quantum 
process called Hawking effect \cite{hawking}. Most of the approaches have been 
developed using a fixed spacetime background 
during the emission process. In this work we implement the idea of Keski-Vakkuri, Kraus and Wilczek \cite{keski,kraus1,kraus2} to the case of a two-dimensional
charged black hole stated earlier \cite{chris}. The energy conservation will 
be the key to our description since it will lead us to the dynamical 
background which according to Parikh and Wilczek \cite{wilczek} may lead to a more realistic description.
The total (ADM) mass will be fixed while 
the mass of the two-dimensional charged black hole will decrease 
due to the emitted radiation. The effect of this modification
in our calculation generalizes in this case our previous results \cite{chris}
which are analogous with the results found in \cite{kwon,parikh,hemming} for the respective geometries;  
a nonthermal partener (``greybody factor"\cite{ofer}) to the thermal spectrum of the Hawking radiation will show up. 
We explore the consequences to the extremal two-dimensional black hole. 
The extremality now will be shifted since the charge $Q$ of the 
two-dimensional charged black hole will be reached by the mass $M$ earlier. 
This alteration produces a non -``frozen" extremal two-dimensional charged black hole characterised by a constant temperature 
$T^{extremal}_{b-h}\neq0$ .  
\section*{Charged Black Hole}
The line element of the charged two-dimensional
black hole is given by (in coordinates corresponding
to the ``Schwarzschild" gauge) :

\be
ds^2 = -g(r)dt^2 + g^{-1}(r)dr^2
\label{linelement}   
\ee
where
\be
g(r) = 1 - \frac{M}{\lambda} e^{-2\lambda r} +
\frac{Q^2}{4 \lambda ^2} e^{-4\lambda r}
\ee
with $0<t<+\infty$, $r_+<r<+\infty$, $r_+$ being the
future event horizon of the black hole. 
\newline
Following a parametrization analogous to the four-dimensional case the metric function factorizes as :
\be
g(r)=(1-\rho_- e^{-2\lambda r})(1-\rho_+e^{-2\lambda r})
\ee
where
\be
\rho_\pm=\frac{M}{2 \lambda}\pm\frac{1}{2\lambda}
\sqrt{M^2-Q^2}
\ee
we can recognize immediately the outer event horizon 
$H^+$ placed at the point $r_+=\frac{1}{2\lambda}ln\rho_+$,
while the ``inner" horizon $H^-$ is at the point
$r_-=\frac{1}{2\lambda}ln\rho_-$. In the extremal case ($Q=M$)
the two surfaces coincide in a single event horizon
at the point :
\be
r_H=\frac{1}{2 \lambda}ln \left( \frac{M}{2 \lambda} \right).
\ee
To deal with phenomena whose main contributions come from the horizon
we must make a coordinate tranformation in order that the new 
coordinates be non-singular on the horizon. We choose the Painlev$\acute{e}$
coordinates \cite{painleve}. We introduce the time coordinate $\tau_P$ :
\bea
\tau_P\hspace{0.05cm} =\hspace{0.05cm} t&&  \nn\\ 
+&\frac{e^{-2\lambda r}\left(\frac{Q^2}{4\lambda^2}-\frac{M}{\lambda}e^{2\lambda r}+e^{4\lambda r}\right)\sqrt{\frac{M}{\lambda}e^{-2\lambda r}-\frac{Q^2}{4\lambda^2}e^{-4\lambda r}}\left(Q^2 -M^2-2M\sqrt{M^2-Q^2}   \right)tanh^{-1}\left({\frac{\sqrt{\frac{M}{\lambda}e^{2\lambda r}-\frac{Q^2}{4\lambda^2}}}{\sqrt{\frac{M^2}{2\lambda^2}-\frac{Q^2}{4\lambda^2}+
\frac{M}{2\lambda^2}\sqrt{M^2-Q^2}}}}\right)}
{\sqrt{\frac{M}{\lambda}e^{2\lambda r}-\frac{Q^2}{4\lambda^2}}
\sqrt{M^2-Q^2}\left(1-\frac{M}{\lambda}e^{-2\lambda r}+\frac{Q^2}{4\lambda^2}     e^{-4\lambda r}\right) 
\sqrt{\frac{M^2}{2\lambda^2}-\frac{Q^2}{4\lambda^2}+
\frac{M}{2\lambda^2}\sqrt{M^2-Q^2}}}&\nn\\
+&\frac{e^{-2\lambda r}\left(\frac{Q^2}{4\lambda^2}-\frac{M}{\lambda}e^{2\lambda r}+e^{4\lambda r}  \right)\sqrt{\frac{M}{\lambda}e^{-2\lambda r}-\frac{Q^2}{4\lambda^2}e^{-4\lambda r}}\sqrt{2M^2-Q^2-2M\sqrt{M^2-Q^2}}\tanh^{-1}\left({\frac{\sqrt{\frac{M}{\lambda}e^{2\lambda r}-\frac{Q^2}{4\lambda^2}}}{\sqrt{\frac{M^2}{2\lambda^2}-\frac{Q^2}{4\lambda^2}-\frac{M}{\lambda^2}\sqrt{M^2-Q^2}}}}\right)}
{2\lambda\sqrt{\frac{M}{\lambda}e^{2\lambda r}-\frac{Q^2}{4\lambda^2}}\sqrt{M^2-Q^2}\left(1-\frac{M}{\lambda}e^{-2\lambda r}+\frac{Q^2}{4\lambda^2}e^{-4\lambda r} \right)}&.
\eea
The line element (\ref{linelement}) is written as :
\be
ds^2 = -(1-\frac{M}{\lambda} e^{-2\lambda r} +
\frac{Q^2}{4 \lambda ^2}e^{-4\lambda r})d\tau_P^2 
 + 2\sqrt{\frac{M}{\lambda} e^{-2\lambda r}-\frac{Q^2}{4 \lambda ^2} e^{-4\lambda r}} d\tau_Pdr + dr^2.
\ee
It is obvious from the above expression that there is no singularity
at the points  $r_+$ and $r_-$. The radial null geodesics followed by
a massless particle are :
\be
\dot{r}\equiv\frac{dr}{d\tau_P}=\pm1-\sqrt{\frac{M}{\lambda} e^{-2\lambda r}-\frac{Q^2}{4 \lambda ^2} e^{-4\lambda r}}
\ee
where the signs $+$ and $-$ correspond to the outgoing and ingoing geodesics,
respectively, under the assumption that $\tau_P$ increases towards future.
Assuming that the total mass is fixed, the mass $M$ of the black hole decreases
to $M-\omega$, where $\omega$ is the shell of energy travelling outwards on the modified geodesics :
\be
\dot{r}=1-\sqrt{\frac{(M-\omega)}{\lambda} e^{-2\lambda r}-\frac{Q^2}{4 \lambda ^2} e^{-4\lambda r}}
\label{geodesic}
\ee 
produced by the modified line element :
\bea
ds^2\hspace{0.2cm} = &\hspace{0.2cm} -\hspace{0.2cm}(1-\frac{(M-\omega)}
{\lambda} e^{-2\lambda r} +
\frac{Q^2}{4 \lambda ^2}e^{-4\lambda r})d\tau_P^2 & \nn\\
&+\hspace{0.1cm} 2\sqrt{\frac{(M-\omega)}{\lambda} 
e^{-2\lambda r}-\frac{Q^2}{4 \lambda ^2} e^{-4\lambda r}} 
d\tau_Pdr 
&\hspace{0.2cm}+\hspace{0.2cm}dr^2.
\eea
It is known that the emission rate from a radiating source can be expressed 
with respect to the imaginary part of the action for an outgoing 
positive-energy particle as :
\be
\Gamma=e^{-2ImS}
\ee
but also with respect to the temperatute and the entropy of the radiating 
source which in our case will be a two-dimensional charged black hole :
\be
\Gamma=e^{-\beta \omega}=e^{-\Delta S_{\hspace{0.05cm}black\hspace{0.1cm}hole}}
\label{temp1}
\ee
where $\beta$ is the inverse temperature of the black hole and 
$\Delta S_{\hspace{0.05cm}black\hspace{0.1cm}hole}$ is the
change in the entropy of the black hole before and after the emission of 
the shell of energy $\omega$ (outgoing massless particle).
It is clear that if we evaluate the action then we will know the temperature 
and/or the change in the entropy of the black hole. We therefore evaluate the 
imaginary part of the action for an outgoing positive-energy particle which crosses the horizon outwards from :
\be
r_{in}=\frac{1}{2\lambda}ln\left[\frac{M}{2 \lambda}+\frac{1}{2\lambda}
\sqrt{M^2-Q^2}\right]
\ee
to 
\be
r_{out}=\frac{1}{2\lambda}ln\left[\frac{(M-\omega)}
{2 \lambda}+\frac{1}{2\lambda}\sqrt{(M-\omega)^2-Q^2}\right].
\ee
The imaginary part of the action will be :
\be
ImS=Im\int_{r_{in}}^{r_{out}}p_{r}dr=Im\int^{r_{out}}_{r_{in}}
\int_{0}^{p_{r}}dp'_{r}dr.
\ee
We make the transition from the momentum variable to the energy variable 
using the Hamilton's equation  $\dot{r}=\frac{dH}{dp_{r}}$  and 
equation (\ref{geodesic}). The result is : 
\be
ImS=Im\int^{r_{out}}_{r_{in}}\int^{\omega}_{0}\frac{(-d\omega')dr}
{1-\sqrt{\frac{(M-\omega')}{\lambda} e^{-2\lambda r}-\frac{Q^2}{4 \lambda ^2} e^{-4\lambda r}}}
\ee
where the minus sign is due to the Hamiltonian being equal to the modified mass $H=M-\omega$. This is not disturbing since $r_{in}>r_{out}$. After some calculations (involving contour integration into the lower half of $\omega'$ 
plane) we get :
\be
ImS=\frac{\pi}{2\lambda}\left[-\omega+\sqrt{(M-\omega)^2-Q^2}
-\sqrt{M^2-Q^2}\right].
\ee
Apparently the emission rate will depend not only on the mass $M$ 
and charge $Q$ of the two-dimensional charged black hole but also 
on the energy $\omega$ of the emitted massless particle :
\be
\Gamma(\omega,M,Q)=e^{-2ImS}=
exp\bigg[\frac{\pi}{\lambda}\left(-\omega+\sqrt{(M-\omega)^2-Q^2}
-\sqrt{M^2-Q^2}\right)\bigg].
\label{temp2}
\ee
Comparing (\ref{temp1}) and (\ref{temp2}) we deduce that the modified 
temperature (due to the specific modelling of the backreaction effect) is :
\be
T= \frac{\lambda}{\pi}\left[1-\sqrt{(\frac{M}{\omega}-1)^2-
(\frac{Q}{\omega})^2} + \frac{1}{\omega}\sqrt{M^2-Q^2}\right]^{-1}
\ee
We see that there are deviations from the standard results \cite{hawking,chris} derived for a fixed background. 
The temperature of the two-dimensional charged black hole is no 
longer the Hawking temperature. These deviations form the so-called ``greybody-factor" that measures the departure from the pure 
blackbody spectrum and lead to a nonthermal spectrum.
A welcomed but not unexpected result is that if we evaluate to first order in $\omega$ temperature we get the result reached in our previous work for the two-dimensional charged black hole :
\be
 T_H=\frac{\lambda}{2\pi}{\mu}
\ee
where $\mu$ is given by :
\be
\mu=1-\frac{\rho_{-}}{\rho_{+}}.
\ee
This constitutes another evidence for our arguments placed here.
\nk
\par
The extremal two-dimensional charged black hole will be created when :
\be
M-\omega=Q 
\ee
This modification indicates that the mass $M$ of the black hole cannot be less than the charge $Q$ (since $\omega=M-Q >0$) and the temperature of the extremal two-dimensional charged black hole will not be zero :
\be
T^{extremal}_{b-h}=\frac{\lambda}{\pi}\left[\frac{(M-Q)}{(M-Q)+\sqrt{M^2-Q^2}}\right].
\ee    
\section*{Discussion}
In this work, we have viewed the Hawking radiation as a quantun tunneling process.
The self-gravitation of the radiation introduced in \cite{wilczek} was included and this treatment introduced 
a nonthermal part for the spectrum. The temperature of the black hole is no more 
the Hawking temperature and the ``greybody factor" showing up declares explicitly 
the emitted particle's energy dependence. The leading term gives the thermal Boltzmann factor while the higher order terms represent corrections emanating 
from the response of the background geometry to the emission of a quantum. The extremal two-dimensional charged black hole is no more ``frozen" but it carries 
a backgound temperature $T^{extremal}_{bh}$ ensuring the validity of 
the third law of black hole thermodynamics. Therefore it is obvious that we have again one more strong evidence to believe that black holes  constitute  excited states while the extremal black holes correspond to ground states.
\par
It is interesting that for the case of the two-dimensional ``Schwarzschild" 
black hole the above procedure of incorporating the effects of the emittance of a shell of energy yields precisely the result obtained with a fixed background \cite{chris,giddings}. The reason for this lack of ``greybody factor" in the 
``Schwarzschild" case is the non - dependence of the Hawking temperature on black hole mass (in contradistinction to the corresponding four-dimensional case). Of course the complete treatment of the backreaction effects remains to be explored further.     
\section*{Acknowledgements}
I would like to thank T. Christodoulakis, G.A. Giamandis and B.C. Georgalas for helpful discussions and kind comments.

 

 \end{document}